\documentclass[%
 reprint,
 amsmath,amssymb,
 prl,
]{revtex4-1}

\usepackage{graphicx}
\usepackage{dcolumn}
\usepackage{bm}

\begin{document}

\preprint{APS/123-QED}

\title{Cascade replication of dissipative solitons}

\author{Bogdan A. Kochetov$^1$}

\author{Vladimir R. Tuz$^{1,2}$}
\email{tvr@jlu.edu.cn; tvr@rian.kharkov.ua}
\affiliation{$^1$International Center of Future Science, State Key Laboratory on Integrated Optoelectronics, College of Electronic Science and Engineering, Jilin University, \\ 2699 Qianjin St., Changchun 130012, China}
\affiliation{$^2$Institute of Radio Astronomy of National Academy of Sciences of Ukraine, \\ 4, Mystetstv St., Kharkiv 61002, Ukraine} 

\date{\today}

\begin{abstract}
We report a new effect of a cascade replication of dissipative solitons from a single one. It is discussed in the framework of a common model based on the one-dimensional cubic-quintic complex Ginzburg-Landau equation in which an additional linear term is introduced to account the perturbation from a particular potential of externally applied force. The effect manifestation is demonstrated on the light beams propagating through a planar waveguide. The waveguide consists of a nonlinear layer able to guide dissipative solitons, and a magneto-optic substrate. In the waveguide an externally applied force is considered to be an inhomogeneous magnetic field which is induced by modulated electric currents flowing along a set of direct conducting wires adjusted on the top of the waveguide.  
\end{abstract}

\maketitle

Dissipative solitons are stable localized structures existing in nonlinear non-equilibrium systems, and they arise due to balance between incoming and outcoming flows of energy or matter within the structures \cite{Akhmediev_Book1}. In contrast to classical solitons in integrable systems, dissipative solitons exhibit entirely new behavior -- they evolve without any constraints on energy or momentum conservation. The nonconservative nature of dissipative solitons involves bifurcations, complex oscillations and self-organization mechanisms which are considered within a new paradigm of nonlinear dynamical systems over several decades \cite{Akhmediev_Book2}. This paradigm has continuously been developed to explain the appearance and nontrivial evolution of self-organized soliton-like structures in hydrodynamical, optical, condensed matter, and biological systems far from equilibrium \cite{Akhmediev_Book1, Akhmediev_Book2, Liehr_Book}.

A wide range of dissipative solitons existing in systems with inertialess nonlinearity can be effectively described in the framework of the complex Ginzburg-Landau equation (CGLE). This model covers phenomena of diverse nature in optical mode-locked and fiber lasers, semiconductor devises, Bose-Einstein condensates, systems with fluid and electro-convection, chemical reactions, etc. \cite{Aranson_RMP_2002}. Being a non-integrable dynamical system near a subcritical bifurcation and having importance for applications the CGLE has attracted much attention \cite{Fauve_PRL_1990, van_Saarloos_PRL_1990, Cross_RMP_1993, Deissler_PRL_1994}. As a result of intense studies a variety of stable localized solutions of the CGLE has been found. They include different forms of stationary and moving solitons \cite{Fauve_PRL_1990, van_Saarloos_PD_1992, Afanasjev_PRE_1996, Soto-Crespo_PLA_2001}, periodically and quasi-periodically pulsating solitons with simple or more complicated behaviors \cite{Deissler_PRL_1994, Akhmediev_PRE_2001}, chaotic solitons \cite{Deissler_PRL_1994, Akhmediev_PRE_2001}, and exploding solitons, which periodically manifest explosive instabilities returning to their original waveforms after each explosion \cite{Soto-Crespo_PRL_2000, Akhmediev_PRE_2001, Soto-Crespo_PLA_2001, Cundiff_PRL_2002}. Remarkably, all the forms of dissipative solitons exist indefinitely in time as long as parameters of the system remain constant, whereas in some range of system parameters the different forms of dissipative solitons can coexist with each other \cite{Afanasjev_PRE_1996, Akhmediev_PRE_2001, Soto-Crespo_PLA_2001}.

An externally applied force upon dissipative solitons influences their waveforms and can be used to control the soliton formation, evolution and existence. For instance, diffusion-induced turbulence in distributed dynamical systems near a supercritical Hopf bifurcation can be modeled by the CGLE with an additional term accounting for a spatial average of the complex amplitude \cite{Battogtokh_PD_1996} or a gradient force \cite{Xiao_PRL_1998}. In optics, evolution of dissipative solitons in an active bulk medium has been studied in the framework of the two-dimensional CGLE with an umbrella-shaped \cite{Yin_JOSAB_2011} as well as a radial-azimuthal \cite{Liu_OE_2013} potentials. The use of an external magnetic field as a driving force to gain control over propagation of solitary waves through magneto-optic systems is another prominent example \cite{Boardman_1995, Boardman_1997, Boardman_2001, Boardman_2003, Boardman_2005, Boardman_2010}. In fact, significant benefits of utilizing a spatially inhomogeneous external magnetic field to acquire different propagation conditions of light dissipative solitons in magneto-optic waveguides have been demonstrated in \cite{Boardman_Chapter_2005, Boardman_2006}. Moreover, a robust mechanism to perform a selective lateral shift within a group of stable dissipative solitons  propagating through a magneto-optic planar structure has been proposed in \cite{OptLett_2017}.

In the present Letter, having considered the classical electromagnetic field in a particular magneto-optic planar nonlinear waveguide being under an action of the external magnetic field, we report, for the first time to the best of our knowledge, a new remarkable effect of the cascade replication of dissipative solitons. 

\begin{figure}[htbp]
\centering\includegraphics[width=\linewidth]{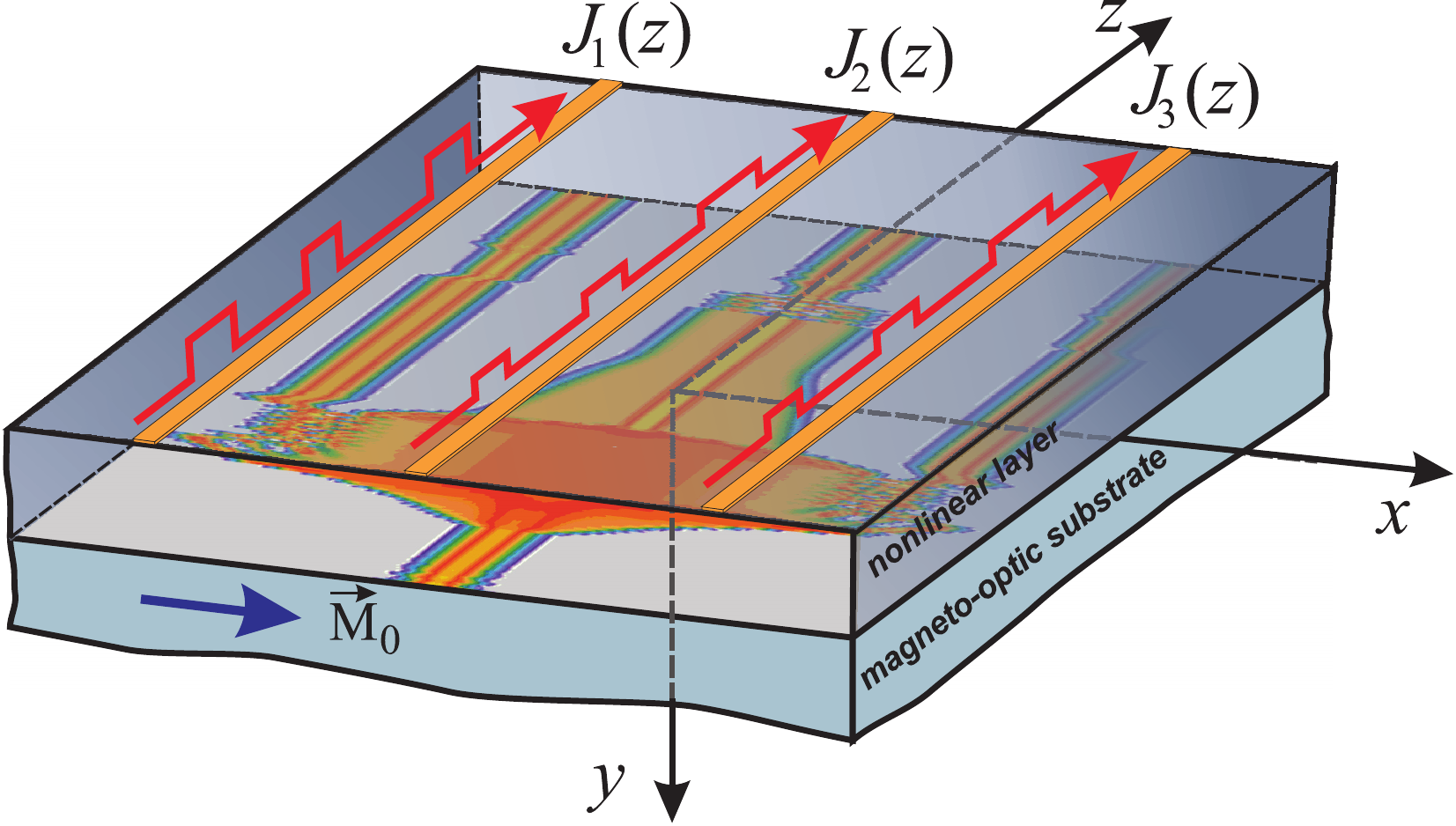}
\caption{Sketch of a planar waveguide. It consists of a nonlinear layer supporting dissipative solitons, and a magneto-optic substrate. Spatially inhomogeneous magnetization $\vec M_0$ is induced by longitudinally modulated electric currents $J_i(z)$ flowing along a set of conducting wires.} \label{fig_1}
\end{figure}

Therefore, further we consider light beams (dissipative solitons) propagating through a planar waveguide, which consists of a nonlinear guiding layer disposed on a magneto-optic substrate (Fig.~\ref{fig_1}). The waveguide is supposed to be infinitely extended along the $x$- and $z$-axes, while along the $y$-axis optical fields are restrained by the boundary conditions on the interfaces of the nonlinear layer. Light beams propagate along the positive direction of the $z$-axis. The external magnetic field is induced by the electric currents $J_i(z)$ that flow through a set of $N$ direct conducting wires (orange straight lines in Fig.~\ref{fig_1}) arranged on the top of the guiding layer. Each current is modulated along the $z$-axis, and its magnitude has a particular piecewise constant profile (zigzag red lines in Fig.~\ref{fig_1}). Thus, the resultant inhomogeneous magnetic field induced by such system of currents acts as an external force which influences upon the dissipative solitions. 

Since the wires are extended along the $z$-axis, the magnetic field component that creates the magnetization is a vector quantity. It is tangential to circles in the ($x$, $y$) plane, centered on each wire. Hence there are components of the magnetization parallel to the $x$- and $y$-axes, but only that parallel to the $x$-axis is important because the components along the $y$-axis give rise to a polar magnetic effect \cite{Boardman_Chapter_2005, Boardman_2006}. The latter causes transverse electric-magnetic (TE-TM) coupling, but since the phase matching condition is not satisfied in this type of guide, the polar effect is negligible.

Therefore, the spatially inhomogeneous distribution of the magnetic field is considered to vary in transverse $x$ and longitudinal $z$ directions only, which is accounted by the scalar magnetization function $Q(x,z)$.  It is expressed in the form \cite{OptLett_2017}
\begin{equation}
\label{Q} Q(x,z) = 2\sum_{i=1}^{N} \tanh\left(\frac{J_i(z)}{47.74\pi\sqrt{(x-x_i)^2+1}}\right),
\end{equation}
where $x_i$ is the $x$ coordinate of the $i$-th current, and the piecewise constant profile function $J_i(z)$ defines the $i$-th current that flows along the corresponding wire.

In the chosen structure geometry the magnetic field influences only the TM modes of the dielectric planar waveguide. Therefore, the evolution of each TM mode within the nonlinear layer can be described in the framework of the following one-dimensional cubic-quintic complex Ginzburg-Landau equation supplemented by the linear potential term \cite{Boardman_Chapter_2005, Boardman_2006}:
\begin{multline}
\label{CQCGLE}
\mathrm{i}\frac{\partial\Psi}{\partial z}+\mathrm{i}\delta\Psi+\left(\frac{1}{2}-\mathrm{i}\beta\right)\frac{\partial^2\Psi}{\partial x^2}+\left(1-\mathrm{i}\varepsilon\right)\left|\Psi\right|^2\Psi \\ - \left(\nu-\mathrm{i}\mu\right)\left|\Psi\right|^4\Psi + Q(x,z)\Psi = 0,
\end{multline}
where the transverse and longitudinal coordinates $x$ and $z$ are normalized on the beam width and the Rayleigh length, respectively, $\Psi\left(x,z\right)$ is a scaled slowly varying electric field envelop, $\delta$ is a linear absorption, $\beta$ is a linear diffusion, $\varepsilon$ is a nonlinear cubic gain, $\nu$ accounts the self-defocusing effect due to the negative sign, and $\mu$ defines quintic nonlinear losses. 

In order to simulate the propagation of dissipative solitons in the planar waveguide under study, we apply the exponential time differentiating method of the second order \cite{Cox_2002} to solve Eq.~\eqref{CQCGLE}. Since the dissipative solitons are objects localized in space we assume that all waveforms arise within the finite interval $[-L_x/2, L_x/2]$. In our numerical calculations, we set $L_x = 80$ and typically sample the computational interval with $2^{12}$ discretization points. The distance along the $z$-axis we sample using the step $10^{-2}$. The parameters of Eq.~\eqref{CQCGLE} are chosen so as to ensure that the so-called `plane pulse' and `composite pulse' solitons are admitted (characteristics of these different pulses, see in \cite{Afanasjev_PRE_1996}). These parameters are: $\beta = 0.5$, $\delta = 0.5$, $\nu = 0.1$, $\mu = 1.0$, and $\varepsilon = 2.52$. In order to launch a stable dissipative soliton into the system, an arbitrary function whose shape is close to the soliton waveform can be used as an initial condition. Thus, in all our calculations we excite the `plane pulse' soliton using the following profile $\Psi(x,0)=\mathrm{sech}(x)$.

The last term in Eq.~\eqref{CQCGLE} accounting for the externally applied force acts as either attractive or repulsive potential corresponding to the positive or negative sign of the function $Q(x,z)$, respectively. Concerning the soliton propagation in magneto-optic waveguides, this action appears as either focusing or defocusing effect influenced by an external magnetic field. Indeed, as pointed out in \cite{Boardman_Chapter_2005, Boardman_2006}, a light beam propagating through a planar magneto-optic waveguide can be focused (defocused) if the beam propagates in the same (opposite) direction as a direct electric current inducing the magnetic field. The focusing effect has already been used to gain a control over the lateral shift of dissipative solitons propagating in the magneto-optic waveguide \cite{OptLett_2017}. Here we employ this nonreciprocal effect to demonstrate a new mechanism of magnetic control over the waveform transitions of dissipative solitons resulting in their cascade replication. 

\begin{figure}[htbp]
\centering
\includegraphics[width=\linewidth]{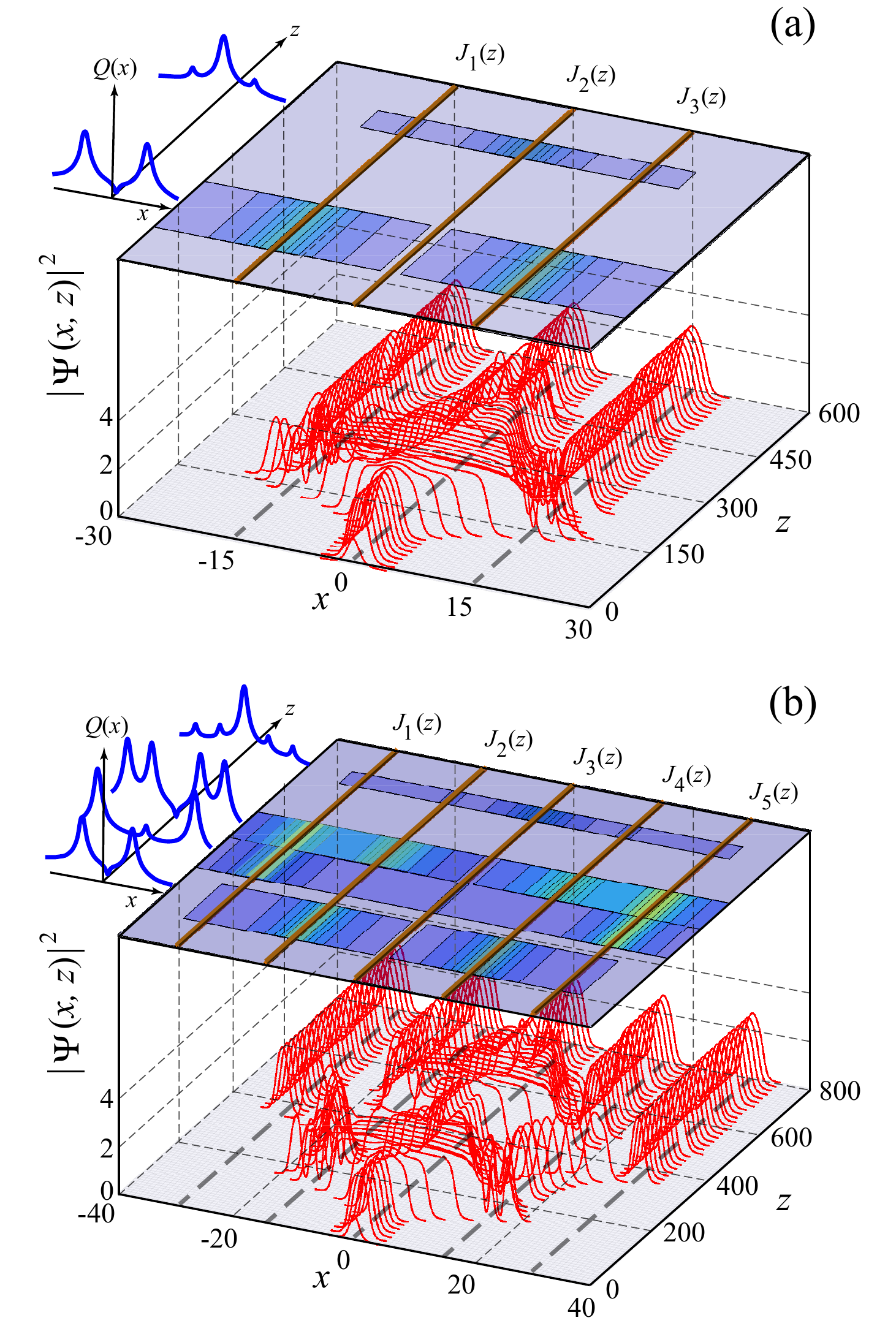}
\caption{Waveform transitions and cascade replication of a dissipative soliton in a magneto-optic planar waveguide; (a) three and (b) five dissipative solitons are produced from a single launched beam due to utilizing the inhomogeneous magnetic field induced by three and five piecewise constant currents modulated along the $z$-axis, respectively. Corresponding parameters of currents, see in Table~\ref{tab1} and Table~\ref{tab2}.}
\label{fig_2}
\end{figure}

Two particular manifestations of such a cascade replication of dissipative solitons in the waveguide structure under study are presented in Fig.~\ref{fig_2} and Supplemental Material
\cite{Suppl_Mat}, where, for instance, three and five light beams are replicated from a single launched beam. The cascade replication appears due to influence of the spatially inhomogeneous magnetic field induced by the electric currents with particular profiles $J_i(z)$. The distribution of the magnetization $Q(x,z)$ is presented in the corresponding color maps situated on the upper edge of each plot, where the conducting wires are drawn as orange straight lines. In the inserts on the left side of each plot the cross-section profiles of this magnetization are depicted with blue solid lines, while in the main parts of the figures red solid lines represent three-dimensional intensity plots of solitons. 

\begin{figure}[htbp]
\centering
\includegraphics[width=\linewidth]{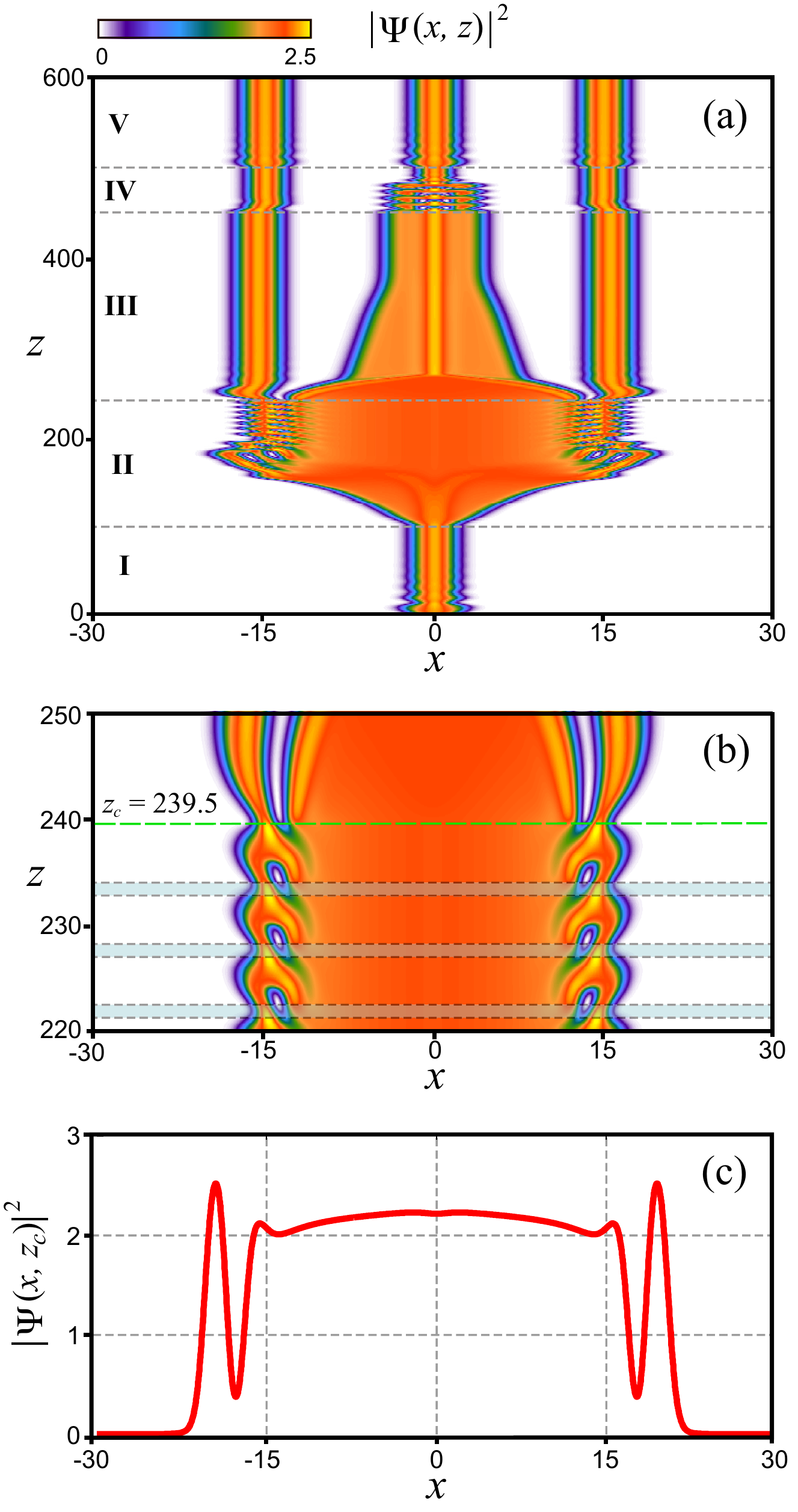}
\caption{Intensity distribution of a dissipative soliton propagating through a magneto-optic planar waveguide; (a) stages of soliton waveform transition; (b) periodical pulsations; areas shaded in light green correspond to zones suitable for replication; green dashed line indicates transition between Stage II and Stage III; (c) waveform of periodically pulsating soliton at the waveguide cross section $z_c=239.5$ that is used for replication. Corresponding parameters of currents, see in Table~\ref{tab1}.}
\label{fig_3}
\end{figure}

In order to explain the particular manifestation of the cascade replication shown in Fig.~\ref{fig_2}(a) overall waveform transitions can be described within five stages. These stages are listed in Table~\ref{tab1} and outlined in Fig.~\ref{fig_3}(a), where they are denoted by Roman numerals. At Stage I, a stable `plain pulse' soliton arises from the initial beam $\Psi(x,0)$ launched into the waveguide. At Stage II three currents with certain magnitudes are switched on to spread the soliton waveform transversely. Along the left and right wires the currents $J_1$ and $J_3$ flow with equal positive magnitudes attracting the soliton, whereas along the middle wire the current $J_2$ has a negative magnitude forming a repulsive potential. Such a current distribution leads to monotonic spreading soliton waveform at early phase, then the short transient mediate phase takes place, which is replaced by the final non-stationary phase, where the soliton waveform changes its profile periodically. The periodical pulsations are disrupted by switching off the currents $J_1$--$J_3$ which initiates Stage III. At Stage III the replication of the single primary soliton into three noninteracting stable ones is fixed. The central soliton has a form of `composite pulse', whereas the remaining two are `plain pulse' stationary solitons. Then, using the focusing mechanism \cite{OptLett_2017}, at Stage IV the central `composite pulse' soliton is transformed into the `plain pulse' soliton. It is achieved by switching on the current of high magnitude to realize the waveform transition of the central beam, whereas the positions and waveforms of the remaining two beams are held by currents of low magnitude. At the final Stage V all currents are switched off, and three exact copies of the primary soliton beam start propagating through the waveguide.

\begin{table}[htbp]
\caption{Parameters of three piecewise constant currents flowing through wires adjusted at positions $\{-15,0,15\}$ on the $x$-axis scale.}
\label{tab1}
\centering
\begin{tabular}{c|c|c}
    \hline
    Stage & Domain, $z$ & Current magnitudes, $J_i(z)$\\
    \hline
    \hline
    I & $[0,100)$ & $\{0, 0, 0\}$\\
    \hline
    II & $[100,239.5)$ & $\{300, -30, 300\}$\\
    \hline
    III & $[239.5,450)$ & $\{0, 0, 0\}$\\
    \hline
    IV & $[450,500)$ & $\{30, 300, 30\}$\\
    \hline
    V & $[500,600)$ & $\{0, 0, 0\}$\\
    \hline
\end{tabular}
\end{table}

In fact, the soliton replication occurs at the certain moment of transition from Stage II to Stage III, when soliton pulsations get broken abruptly by switching off the currents. This important transition between Stage II and Stage III is highlighted in Fig.~\ref{fig_3}(b), where three last periodical waveform pulsations and the start of Stage III are zoomed in. One can see that the periodical pulsations arise and persist as long as the currents magnitudes remain unchanged. Switching off the currents turns the periodically pulsating soliton waveform back to the stationary one. However, depending on the phase of periodical pulsations when the currents are switched off, the soliton has an alternative to turn back either to one `composite pulse' or to three noninteracting solitons. In the latter case two `plane pulse' and one `composite pulse' solitons arise.

Thus, the ability of soliton replication depends critically on the waveform profile that exists within the soliton pulsation period at which the currents need to be switched off. Three periodical zones containing the waveform profiles suitable for replication along the waveguide length are outlined and shaded with light green in Fig.~\ref{fig_3}(b). Within the outlined zones, switching off the currents results in transition of the pulsating soliton into two `plane pulse' and one `composite pulse' solitons which are suitable for subsequent manipulations. Switching off the current outside these zones is inappropriate, since only single `composite pulse' soliton appears in this case. Such an occurrence of two different scenarios is explained by the fact that both the `plane pulse' and `composite pulse' solitons can coexist in the given system under the chosen set of equation parameters. The waveform of pulsating soliton used for replication is presented in Fig.~\ref{fig_3}(c). It is fixed at the waveguide cross section $z_c=239.5$ which is indicated by the green dashed line in Fig.~\ref{fig_3}(b).

The cascade replication of dissipative solitons can be performed repeatedly to generate more light beams within the waveguide. This ability is presented in Fig.~\ref{fig_2}(b), where the cascade replication has been achieved two times in a row. The simulation parameters are summarized in Table~\ref{tab2}. This cascade replication is performed in eight stages. The first three stages I-III repeat the current manipulations described above. Stage IV is intended to shift the replicated `plain pulse' solitons away from the central `composite pulse' soliton. The currents $J_1$ and $J_5$ of high magnitude induce focusing magnetic field that shifts the corresponding solitons, while the  current $J_3$ of low magnitude is switched on to hold the central `composite pulse' soliton on the way of its propagation. The manipulations performed at stages V-VIII just repeat those of the stages II-V for the cascade replication. As a result of these manipulations, five exact copies of the primary beam are obtained in the waveguide.  

\begin{table}[htbp]
\caption{Parameters of five piecewise constant currents flowing through wires adjusted at positions $\{-30,-15,0,15,30\}$ on the $x$-axis scale.}
\label{tab2}
\centering
\begin{tabular}{c|c|c}
    \hline
    Stage & Domain, $z$ & Current magnitudes, $J_i(z)$\\
    \hline
    \hline
    I & $[0,100)$ & $\{0,0,0,0,0\}$\\
    \hline
    II & $[100,239.5)$ & $\{0,300,-30,300,0\}$\\
    \hline
    III & $[239.5,280)$ & $\{0,0,0,0,0\}$\\
    \hline
    IV & $[280,400)$ & $\{300,0,30,0,300\}$\\
    \hline
    V & $[400,503.3)$ & $\{300,300,-30,300,300\}$\\
    \hline
    VI & $[503.3,650)$ & $\{0,0,0,0,0\}$\\
    \hline
    VII & $[650,700)$ & $\{30,30,300,30,30\}$\\
    \hline
    VIII & $[700,800)$ & $\{0,0,0,0,0\}$\\
    \hline
\end{tabular}
\end{table}

In conclusion, we considered the effect of cascade replication of dissipative solitons in a system with inertialess nonlinearity governed by the one-dimensional cubic-quintic CGLE with an additional linear term supplemented to account for the influence of an externally applied force upon the solitons. We found out that the replication is only possible when a periodically pulsating soliton waveform starts the bifurcation transition to a stationary one being within the periodical narrow zones. 

As a practical model, we studied the effect of cascade replication of light dissipative solitons propagating through a magneto-optic planar waveguide, where an inhomogeneous magnetic field plays a role of the external force that influences upon the light beams. As a result of particular manipulations with the modulated currents that induce this magnetic field, three and five exact copies of the primary beam are obtained. 

Since we used the common mathematical model based on the CGLE we predict that  the effect of soliton replication can be achieved in systems of diverse nature.

\bibliography{cascade_solitons}

\end{document}